\def\be{\begin{equation}}
\def\ee{\end{equation}}
\documentstyle[aps,prl]{revtex}
\begin{document}
\twocolumn
[
\draft
\title{Field-induced Commensurate-Incommensurate phase transition
in a Dzyaloshinskii-Moriya spiral antiferromagnet.}
\author{A. Zheludev, S. Maslov, and G. Shirane}
\address{
Department of Physics, Brookhaven National
Laboratory, Upton, New York 11973}
\author{Y. Sasago, N. Koide, and K. Uchinokura}
\address{
Department of Applied Physics, The University of Tokyo,\\
7-3-1 Hongo, Bunkyo-ku, Tokyo 113, Japan}
\date{\today}
\maketitle
\widetext
\advance\leftskip by 57pt
\advance\rightskip by 57pt

\begin{abstract}
We report an observation of a commensurate-incommensurate phase transition
in a Dzyaloshinskii-Moriya spiral magnet Ba$_2$CuGe$_2$O$_7$. The
transition is induced by applying a magnetic field in the plane of spin
rotation. In this experiment we have direct control over the strength of
the commensurate potential, while the preferred incommensurate period of
the spin system remains unchanged.  Experimental results for the period of
the soliton lattice and bulk magnetization as a function of external
magnetic field are in quantitative agreement with theory.
\end{abstract}
\pacs{}
]
\narrowtext

Studies of Commensurate-Incommensurate (CI) phase transitions have a
long history, dating back to pioneering works of Frenkel and Kontorova
\cite{Frenkel38} and Frank and van der Merwe.\cite{Frank49}. Since then
CI transitions were discovered and studied in a number of such seemingly
unrelated systems as noble gas monolayers adsorbed on graphite
surface\cite{Clarke80,Zabel80}, charge density wave
materials\cite{Wilson75,Fleming78}, and rare-earth
magnets\cite{Koehler72} (For comprehensive reviews see for example
Ref.~\cite{Bak82}). As a rule, CI transitions result from a competition
between two distinct terms in the Hamiltonian that have different
``built-in'' spatial periodicities and are often referred to as
potential and elastic energy, respectively. The potential energy by
definition favors a structure commensurate with the crystal lattice. The
elastic term is intrinsic to the system where the transition occurs, and
has a different ``natural'' built-in period. In many known realizations
of CI, such as adsorbed gas monolayers, it is the period set by the
elastic term that can be varied in an experiment to drive the
transition, whereas both the strength and the period of the potential
remain constant. In other systems, such as rare-earth magnets, both the
elastic term (exchange coupling between spins) and the potential
(magnetic anisotropy) can be changed, but only indirectly, by varying
the temperature.

From the very start it was clear that in its purest form a CI transition
may be driven by the change of the {\it strength} of the potential
alone, with the two ``built-in'' periods remaining constant. An elegant
experimental realization of this type of CI was first proposed by
Dzyaloshinskii \cite{Dzyaloshinskii65}, who considered an incommensurate
spiral magnetic structure in a magnetic field applied {\it in the plane
of rotation of spins}. In this model incommensurability is intrinsic to
the spin system and results from spin-spin interactions. The role of the
potential is played solely by an external magnetic field $H$, that
favors a commensurate spin-flop state. In a real magnetic material with
this type of CI transition an experimentalist would have a convenient
handle on the strength of the potential term, adjusting it by simply
changing the magnetic field. Such systems are not easy to find. In most
spiral magnets, e.g., cubic MnSi \cite{Ishikawa76} and FeGe
\cite{Lebech89}, even a small field is sufficient to realign the spin
plane {\it perpendicular} to the field direction. While RbMnBr$_3$
\cite{Zhitomirsky95} may actually be one material where this does not
occur, to our knowledge, the most crucial quantities, namely the
incommensurability parameter $\zeta$ and magnetization $M$, have not
been measured as a function of the external field in any ``clean''
realization of Dzyaloshinskii's model to date. In the present paper we
report the first direct experimental observation of a
Dzyaloshinskii-type field-induced CI transition in
Ba$_{2}$CuGe$_2$O$_7$, also presenting experimental data for $\zeta(H)$
and $M(H)$.  On the theoretical side we go beyond a qualitative
analysis of the critical properties (close to the phase transition), as
was previously done by Dzyaloshinskii. Dealing with this particular
system, we construct an exactly solvable model and derive exact results
that are in {\it quantitative} agreement with experiment throughout the
{\it entire} phase diagram.

The structural and magnetic properties of Ba$_{2}$CuGe$_2$O$_7$ are
discussed in detail in Ref.~\cite{Zheludev96-BACUGEO}. In the layered
tetragonal crystal structure the magnetic Cu$^{2+}$ sites form a square
lattice with nearest-neighbor (nn) distances of 6~\AA~along the (1,1,0)
and (1,-1,0) directions, respectively. The low-temperature
($T_{N}=3.26$~K) magnetic phase is a weak distortion of the N\'{e}el
spin arrangement, with all spins confined to the $(1,-1,0)$ plane and
the staggered magnetization slowly rotating upon translation along the
$(1,1,0)$ direction. The propagation vector is $(1+\zeta,\zeta,0)$,
where $\zeta=0.027$. Only the nearest-neighbor in-plane
antiferromagnetic exchange constant is significant and is equal to
$J_{ab}=0.48$~meV. The coupling between adjacent Cu-planes is
ferromagnetic, with $|J_c| /|J_{ab}| \approx 1/37$. We have previously
suggestedthat the incommensurate structure is a result of
Dzyaloshinskii-Moriya (DM) \cite{Dzyaloshinskii57,Moriya60}
antisymmetric exchange interactions. The corresponding term in the
Hamiltonian can be written as ${\bf D} ({\bf S}_1 \times {\bf S}_2)$,
where $\times$ denotes the vector product and ${\bf D}$ is a vector
associated with the oriented bond between the two interacting spins.

From the symmetry properties of the lattice we deduce that for a bond
between two nn Cu sites along the $(1,1,0)$ direction, the only
allowed components of ${\bf D}$ are those along $(1,-1,0)$ and $(0,0,1)$,
respectively (Fig. \ref{layer}). The $(1,-1,0)$-component does not change
sign from one bond to the next, while the $(0,0,1)$ component is sign-alternating. It
is the uniform component that is liable for the incommensurate distortion  of
the N\'{e}el structure, and for the rest of the paper we shall ignore the
oscillating component, assuming that ${\bf D}$ is in the $(a,b)$ plane. The
interaction energy is minimized when all spins are perpendicular to ${\bf
D}$. The total exchange energy of the pair of nearest-neighbor spins is
given by $2J_{ab} \cos \phi +D \sin \phi=\sqrt{ 4J_{ab}^2 +D^2} \cos(\phi
-\alpha)$, where $\phi$ is the angle between spins, and $\alpha =
\arctan D/2J_{ab}$. The energy is a minimum at $\phi=\pi+\alpha$. The classical
ground state is therefore a spin-spiral, with all spins in the (1,-1,0)
plane, and the angle between subsequent spins equal to $\pi+\alpha$. The
angle $\alpha$ is related to the propagation vector $\zeta$ by $\alpha=2
\pi \zeta\approx 10^\circ$. Precisely this spin configuration was found in
the initial zero-field neutron diffraction experiments 
\cite{Zheludev96-BACUGEO}.

The central experimental result of this paper is the observation of a CI
transition in Ba$_{2}$CuGe$_2$O$_7$, induced by a magnetic field applied
along the $(0,0,1)$ direction. Single-crystal magnetization measurements
were performed using a conventional DC-squid magnetometer in the
temperature range 2--300~K. Experimental $\chi(H)\equiv \case{dM}{dH}$
for $T=2$~K are shown in Fig.~\ref{mag}. For ${\bf H}||{\bf a}$ no
anomalies are observed. In contrast, when the field is applied along the
$c$ axis, a distinct feature is seen around $H=2$~T and indicates the
presence of a magnetic phase transition. Neutron diffraction experiments
were carried out on the H9 (cold beam) and H4M (thermal beam) 3-axis
spectrometers at the High Flux Beam Reactor (HFBR) at Brookhaven
National Laboratory on a $\approx 4\times 4\times4$~mm$^{3}$
single-crystal sample, in the temperature range 1.3--5~K and magnetic
fields up to 6.5~T, applied along the $c$-axis of the
crystal\cite{elsewhere}. In zero field, at $T=2.4$~K$<T_{N}$, elastic
scans along the $(1+\zeta,\zeta,0)$ direction show magnetic Bragg
reflections centered at an incommensurate position $\zeta=\pm 0.0273$
(Ref.~\cite{Zheludev96-BACUGEO}, Fig. 3b). As the magnetic field
increases, the peak moves in closer to the antiferromagnetic zone-center
at $(1,0,0)$ [Fig.~\ref{neutron}, insert (a)]. At $H>H_{c}\approx 2.3$~T
the satellites at $(1\pm \zeta,\pm \zeta,0)$ are no longer observed, but
are replaced by a single peak at the C-point $(1,0,0)$ [Fig.~\ref{neutron},
insert (b)]. The magnetic structure thus becomes commensurate. The
experimental field-dependence of the propagation vector $\zeta$ is shown
in the main panel of Fig.~\ref{neutron}.

To quantitatively describe the field-dependent behaviour, we follow the
approach of Dzyaloshinskii \cite{Dzyaloshinskii65}. We assume that the
vector of local staggered magnetization at point ${\bf r}$ remains in the
$(1,-1,0)$ plane and forms angle $\theta ({\bf r})$ with respect to the
$c$-axis. The free energy per Cu-plane in the continuous limit is then given by:
\begin{eqnarray}
F&=& \int  [
{\rho _s \over 2}
\left( \left({\partial \theta ({\bf r}) \over
\partial x} 
-{\alpha \over
\Lambda}\right)^2+\left({\partial \theta ({\bf r}) \over
\partial y}\right)^2 + \gamma \left( {\partial \theta ({\bf r}) \over
\partial z} \right)^2 \right) - \nonumber \\
&-& {(\chi_{\perp}-\chi_{\|}) H^2 \over 2}
 \sin ^2 \theta ({\bf r}) ] d{\bf x} d{\bf y}\qquad.
\label{energy}
\end{eqnarray}
Here the axes $x$, $y$ and $z$ run along the $(1,1,0)$, $(1,-1,0)$ and
$(0,0,1)$ directions, respectively, and ${\bf r}=(x,y,z)$.
The first term is the total elastic energy of exchange interactions, and
favors a spiral structure of period $\case{2\pi \Lambda}{\alpha}$. In
Eq. (\ref{energy}) $\Lambda$ is the in-plane nn Cu-Cu distance, and $\rho
_s$ is the in-plane spin stiffness, that for classical spins at zero
temperature is given by $\rho_s (0)= \sqrt { J_{ab}^2 +D^2}\ S^2$. $\gamma$
is the spin stiffness anisotropy defined by $\gamma (\Lambda_c/ \Lambda)^2
= |J_c|/|J_{ab}| \approx 1/37$ for Ba$_2$CuGe$_2$O$_7$. The second term
represents the Zeeman energy. $\chi_{\|}(T)$ and $\chi_{\perp}(T)$ are
defined as magnetic susceptibilities with respect to fields that rotate
along with the spiral structure and are parallel or perpendicular to the
local staggered magnetization, respectively. In the paramegnetic phase
$\chi_{\|}(T)=\chi_{\perp}(T)$, while at $T=0$ the classical result is
$\chi_{\perp} (0)= (g \mu_B)^2/(8 J_{ab}\Lambda ^2 )$, and $\chi_{\|}
(0)=0$. The equilibrium spin configuration should minimize the free energy
(\ref{energy}), and therefore satisfy
\begin{equation}
 {\partial ^2 \theta \over
\partial x^2}=-{(\chi_{\perp}-\chi_{\|}) H^2 \over \rho _s}
\sin \theta \cos \theta = -{1 \over 2 \Gamma^2} \sin 2\theta \qquad ,
\label{sg}
\end{equation}
where $\Gamma=(\rho _s/H^2 (\chi_{\perp}-\chi_{\|}))^{1/2}$.

Expression (\ref{sg}) has the form of the sine-Gordon equation,
that is central to describing CI transitions in many systems, and
its ``soliton lattice'' solutions are well-known:
\begin{equation}
\theta (x)=am(x/ \beta \Gamma, \beta),  \qquad \label{solution}
\end{equation}
where $am(x, \beta)$ is the Jacobi elliptic function of modulus $\beta$.
Analogues of Eqs. (\ref{energy}) and (\ref{sg}) were derived in Ref.
\cite{Dzyaloshinskii65}. To obtain exact results for $\zeta(H)$ and $M(H)$, however, we
make one additional crucial step. For each value of $H$ of all valid
solutions, labeled by $\beta$, one has to choose the one that indeed
corresponds to the global minimum of the free energy. This is done by
substituting Eq. (\ref{solution}) into Eq. (\ref{energy}) and minimizing
the resulting expression with respect to $\beta$. After some algebra we find
that $F$ is minimized when
\begin{eqnarray}
{\beta \over E( \beta)}={H \over H_c} \qquad ; \label{az-2}\\
H_c= {\pi \alpha \over 2 \Lambda} \sqrt{\rho_s \over \chi_{\perp}-\chi_{\|}}
\qquad,
\label{az-1}
\end{eqnarray}
where $E(\beta)$ is the elliptic integral of the second kind. $H_{c}$ is
the critical field at which the CI transition occurs
\cite{Dzyaloshinskii65}. Indeed, for $H>H_{c}$ the spin structure is
given by the soliton-free solution $\theta({\bf r})\equiv \pi /2$, which
corresponds to a commensurate spin-flop phase, as visualized in
Fig.~\ref{spins}(a). The staggered magnetization in this case is
parallel to the $x$-axis and the spins are slightly tilted in the
direction of the field. In the limit $H=0$ one has $\beta \rightarrow 0$
and $\theta(x)=\case{\alpha}{2 \pi} \case{x}{\Lambda}$, which
corresponds to an unperturbed sinusoidal spin-spiral
[Fig.~\ref{spins}(c)]. Most interesting is the case $0<H<H_{c}$, where
the spin structure may be described as a soliton lattice: regions of the
spin-flop phase are interrupted at regular intervals by magnetic domain
walls, or solitons [Fig.~\ref{spins}(b)]. In each soliton the direction
of staggered magnetization rotates by an angle $\pi$. At $H\rightarrow
H_{c}$, the density of solitons starts to decrease very rapidly, as
$1/|\ln (H_c-H)| $\cite{Dzyaloshinskii65}. The transition is thus almost
first order: as $H\rightarrow H_{c}$, the two magnetic satellites at $(1\pm
\zeta,\pm \zeta,0)$ converge to eventually produce a single peak at
$(1,0,0)$.

The exact expression for $\beta(H)$ [Eq.~(\ref{az-2})] enables us to derive parametric
equations for the field-dependence of magnetization and incommensurability
parameter, and directly compare these predictions to experimental results
for Ba$_{2}$CuGe$_2$O$_7$. Using the formula $M=-\partial F/ \partial H$ and the
equalities for derivatives of elliptic functions \cite{Ryzhik} one gets:
\begin{equation}
M=\chi_{\|} H + (\chi_{\perp}-\chi_{\|}) H
\frac{1}{\beta^{2}}\left(1-\frac{E(\beta)}{K(\beta)}\right) \qquad. \label{magn}
\end{equation}
The magnetization curve is continuous at the critical field, while
$\chi(H)\equiv \case{dM}{dH}$ diverges as $H_c$
is approached from below, and is constant and equal to $\chi_{\bot}$ at
$H>H_{c}$. In zero field $\chi$ is equal to $(\chi_{\|}+\chi_{\perp})/2$.
We now use Eq. (\ref{magn}) together with the formula (\ref{az-2}) to fit
the experimental $\chi(H)$ for Ba$_{2}$CuGe$_2$O$_7$ measured at $T=2$~K. With
$\chi_{\bot}=3.43 \times 10^{-5}$~emu/g, $\chi_{||}=0.89 \cdot
10^{-5}$~emu/g and $H_{c}=1.88$~T a very good fit is obtained
(Fig.~\ref{mag}, solid line for $H||c$). In the refinement we have included
a linear contribution to $\chi(H)$. This term is present in both
$\chi_{c}(H)$ and $\chi_{a}(H)$ and is a result of non-linear corrections
to the local susceptibility, neglected in our treatment.

As $H\rightarrow H_{c}$ the soliton density decreases, i.e., the period
of the magnetic structure increases with increasing field. In our case
expressing $\zeta$ in terms of $\beta$ is rather straightforward and
yields:
\begin{equation}
{\zeta(H) \over \zeta(0)}={H \over H_c} {\pi ^2 \over 4
\beta  K( \beta)}={\pi^2 \over 4 E(\beta) K(\beta)} \qquad .
\label{zeta}
\end{equation}
Again, Eq. (\ref{zeta}), when combined with Eq. (\ref{az-2}), gives a
parametric curve for $\zeta(H)$, that can be fit to the experimental
data for Ba$_{2}$CuGe$_2$O$_7$ using $H_{c}$ as the only adjustable
parameter [$\zeta(0)=0.027$ is measured independently]. The result with
$H_{c}=2.13$~T is shown in a solid line in Fig.~\ref{neutron}. A
remarkable agreement is obtained except very close to $H_{c}$.
Discrepancies close to the transition point are to be expected, since
the transition is almost first order, the soliton lattice is very soft
and easily pinned by any impurities. The same effects prevent us from
experimentally observing a true divergence in $\chi(H)$ at the critical
field (Fig.~\ref{mag}).

Finally, we can  estimate the actual value of the critical field
using the previously measured exchange constants and $\zeta(0)$. With
$J_{ab}=0.48$~meV\cite{Zheludev96-BACUGEO} the exchange energy per bond is
$\tilde{J}=2J_{ab}S^{2}\approx0.24$~meV. For $\chi_{||}$ and $\chi_{\bot}$ we
can use the values for a classical Heisenberg antiferromagnet at $T=0$. ESR
measurements\cite{Sasago-ESR} provide us with the gyromagnetic ratios of
Cu$^{2+}$: $g_{a}=2.044$ and $g_{c}= 2.474$. Using Eq. (\ref{az-1}) and
$\chi_{\bot}=(g\mu_{B})^{2}/(8 J)$ we
obtain $H_{c}=3.3$~T, that should be compared to the experimental value
$H_{c}\approx 2.1$~T. Considering that in these estimates we
have completely ignored quantum- and temperature corrections to $\chi$ and
$\rho_{s}$, a 30\% consistency is indeed acceptable.

In summary, we have observed a rare type of CI transition that is driven
exclusively by the changing strength of the commensurate potential. The
latter is directly controlled in an experiment by varying the magnetic
field. A transition of this kind was envisioned over three decades ago
by Dzyaloshinskii, and now we find that Ba$_{2}$CuGe$_2$O$_7$ exhibits
it in its original form. Now that the underlying physics  is rather well
understood, Ba$_{2}$CuGe$_2$O$_7$ can be used as a very neat and simple
model system for further studies of spiral magnetism. For example, it
would be very interesting to look closer at the magnetic critical
behavior. Another promising direction for future work is the study of
the effect of an in-plane magnetic field.

We wish to thank P. Bak, V. Emery and I. Zaliznyak for useful discussions.
This study was supported in part by NEDO (New Energy and Industrial
Technology Development Organization) International Joint Research Grant and
the U.S. -Japan Cooperative Program on Neutron Scattering. Work at
Brookhaven National Laboratory was supported by the U.S. Department of
Energy Division of Material Science, under contract DE-AC02-76CH00016.


\begin{thebibliography}{10}
\bibitem{elsewhere}
A complete experimental report will be published elsewhere.
\end{thebibliography}

\begin{figure}
\caption{A schematic view of a Cu-Ge-O layer in Ba$_2$CuGe$_2$O$_7$. The
arrows indicate the components of Dzyaloshinskii vector $\vec{D}$, allowed by
the symmetry.}
\label{layer}
\end{figure}

\begin{figure}
\caption{Field dependence of the magnetic susceptibility measured in Ba$_2$CuGe$_2$O$_7$
at $T=2$~K for the magnetic field applied along the $c$ (circles) and
$a$ (crosses) axes of the crystal, respectively. The solid line is a
theoretical fit to the data, as described in the text.}
\label{mag}
\end{figure}

\begin{figure}
\caption{Field dependence of the magnetic propagation vector in Ba$_2$CuGe$_2$O$_7$ measured
at $T=2.4$~K. The solid line is a theoretical fit given by Eqs.
(\protect\ref{az-2},\protect\ref{zeta}). Insert: Elastic scans across the
antiferromagnetic zone center measured in Ba$_2$CuGe$_2$O$_7$ at $T=2.4K$
for two different values of magnetic field applied along the $(0,0,1)$
direction.}
\label{neutron}
\end{figure}

\begin{figure}
\caption{Spin configurations for  the spin-flop phase at $H>H_c$ (a),
the soliton lattice at $H=0.997 H_c$ (b) and the circular spin spiral at $H=0$ (c).}
\label{spins}
\end{figure}

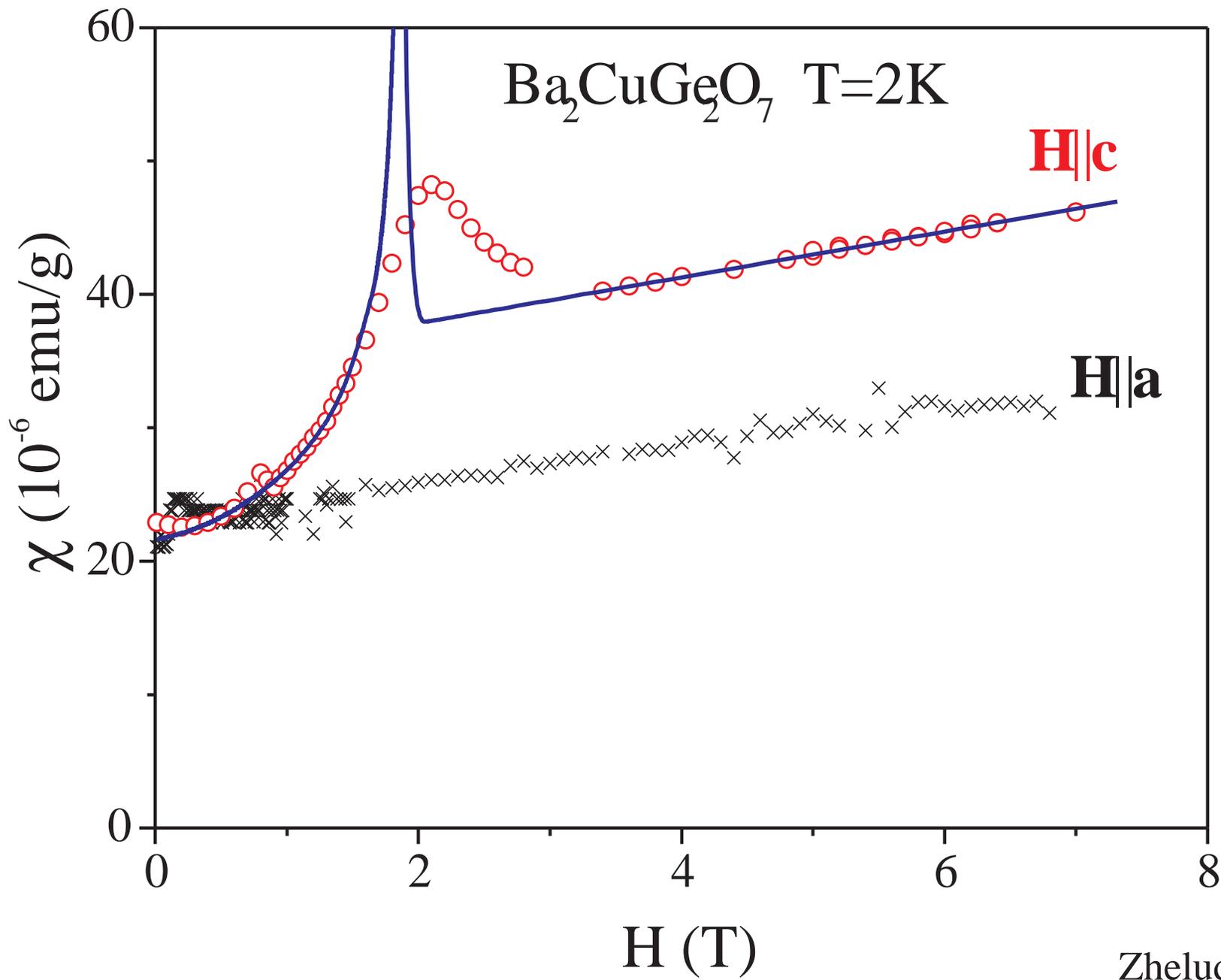
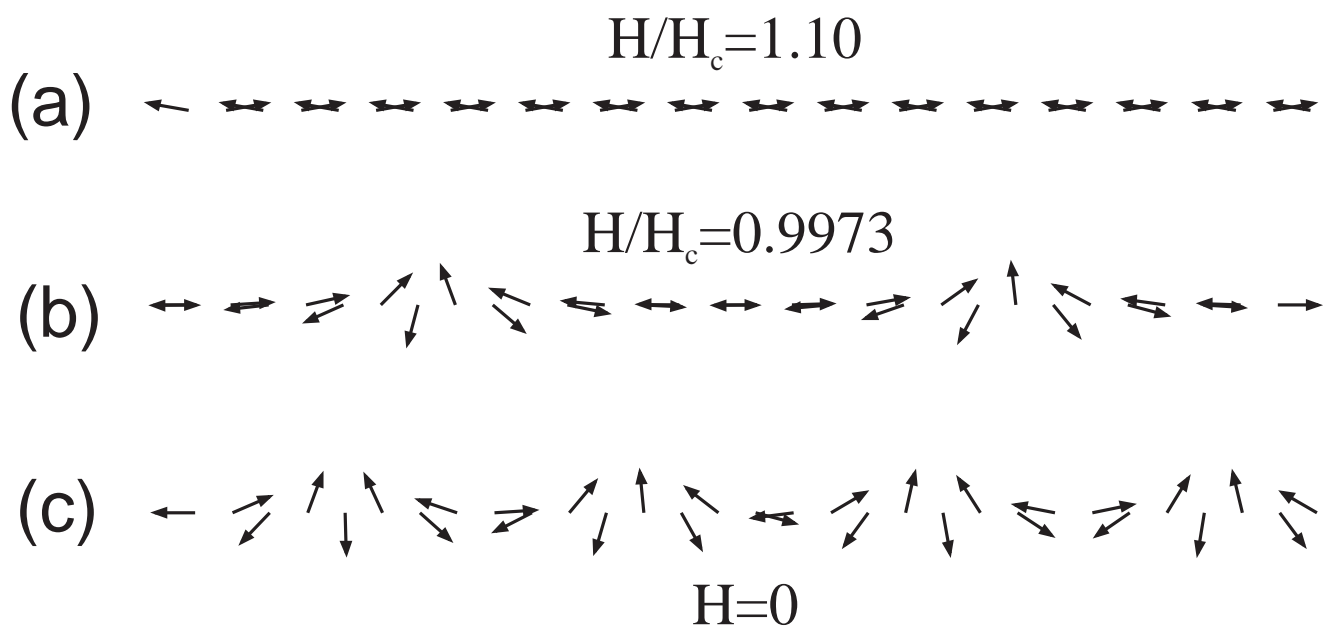
\end{document}